\documentclass[twocolumn,superscriptaddress,prl]{revtex4}    
\usepackage{graphicx,amsmath,mathrsfs}
\usepackage{amssymb}
\usepackage{amsthm,multirow}
\usepackage{bm,bbm}
\usepackage{subfigure}   

\usepackage{color}

\def\bc{\begin{center}}
\def\ec{\end{center}}

\newcommand{\up}{\uparrow}
\newcommand{\dw}{\downarrow}
\newcommand{\pd}{{\phantom{\dagger}}}

\newcommand{\CC}{{\cal C}}

\def\ie{\emph{i.e.},\ }
\def\eg{\emph{e.g.}\ }

\begin{document}
\title{Quantized charge transport in chiral Majorana edge modes}
\author{Stephan Rachel}
\affiliation{Institut f\"ur Theoretische Physik, Technische Universit\"at Dresden, 01062 Dresden, Germany}
\author{Eric Mascot}
\affiliation{Department of Physics, University of Illinois at Chicago, Chicago, IL 60607, USA}
\author{Sagen Cocklin}
\affiliation{Department of Physics, University of Illinois at Chicago, Chicago, IL 60607, USA}
\author{Matthias Vojta}
\affiliation{Institut f\"ur Theoretische Physik, Technische Universit\"at Dresden, 01062 Dresden, Germany}
\affiliation{Center for Transport and Devices of Emergent Materials, Technische Universit\"at Dresden,
01062 Dresden, Germany}

\author{Dirk K.\ Morr}
\affiliation{Department of Physics, University of Illinois at Chicago, Chicago, IL 60607, USA}
\date{\today}

\maketitle

%
%

\textbf{
Majorana fermions\,\cite{majorana37nc171} can be realized as quasiparticles in topological superconductors\,\cite{lutchyn-10prl077001,oreg-10prl177002,mourik-12s1003}, with potential applications in topological quantum computing\,\cite{moore-91npb362,read-00prb10267,ivanov01prl268,nayak-08rmp1083}. Recently, lattices of magnetic adatoms deposited on the surface of $s$-wave superconductors --  Shiba lattices -- have been proposed as a new platform for topological superconductivity\,\cite{li-16nc12297,rontynen-15prl236803}. These systems possess the great advantage that they are accessible via scanning-probe techniques, and thus enable the local manipulation and detection of Majorana modes. Realizing quantum bits on this basis will rely on the creation of nanoscopic Shiba lattices, so-called Shiba islands. Here, we demonstrate that the topological Majorana edge modes of such islands display universal electronic and transport properties. Most remarkably, these Majorana modes possess a quantized charge conductance that is proportional to the topological Chern number, $\CC$, and carry a supercurrent whose chirality reflects the sign of $\CC$. These results establish nanoscopic Shiba islands as promising components in future topology-based devices.
}

\bigskip

Originally proposed as elementary particles with the peculiar property of being their own antiparticles\,\cite{majorana37nc171},
Majorana fermions have been argued
to exist in several condensed-matter settings: in certain fractional quantum Hall states\,\cite{moore-91npb362}, in superfluid He-3\,\cite{volovik88zetf123}, in fractionalized spin liquids \cite{kitaev06}, in chiral superconductors\,\cite{kallin-16rpp054502}, and in hybrid superconducting structures \cite{mourik-12s1003,nadj-perge-14s602,ruby-15prl197204,pawlak-16npj16035}. Besides being of fundamental interest as exotic particles, Majorana fermions hold great potential for realizing fault-tolerant topological quantum bits due to their non-Abelian braiding statistics\,\cite{moore-91npb362,read-00prb10267,ivanov01prl268,nayak-08rmp1083}. Realizing this potential, however, necessitates the ability to create, detect, and manipulate single Majorana fermions in nanoscopic topological superconductors.

First steps towards achieving this goal have been taken through the creation of Shiba chains -- chains of magnetic Fe atoms -- on the surface of Pb, an $s$-wave superconductor with strong Rashba spin-orbit interaction\,\cite{li-14prb235433,nadj-perge-14s602,ruby-15prl197204,pawlak-16npj16035}. These chains were shown to realize one-dimensional topological superconductivity, with Majorana zero modes being localized at the chain ends\,\cite{nadj-perge-14s602,ruby-15prl197204,pawlak-16npj16035}. Such surface systems can be investigated via scanning tunneling spectroscopy (STS), providing not only spectroscopic but also spatially resolved insight into the nature of Majorana modes. Most recently, the concept of Shiba chains was theoretically generalized to Shiba lattices\,\cite{li-16nc12297,rontynen-15prl236803} -- two-dimensional lattices of magnetic adatoms placed on an $s$-wave superconductor -- as a new platform for chiral superconductivity with dispersing Majorana edge modes. Such systems are characterized by a non-zero Chern number $\CC$, similar to the integer quantum Hall effect, and display a universal thermal Hall conductivity $\kappa^{xy}/T = \CC \frac{\pi^2 k_B^2}{3h}$\,\cite{chiu-16rmp35005}. Whether the Majorana edge modes in Shiba lattices also exhibit universal {\it charge} transport properties, in analogy to the quantized charge Hall conductivity $\sigma^{xy} = \frac{e^2}{h} \CC$ in quantum Hall systems \,\cite{thouless-82prl405}, that would allow one to uniquely identify them, is presently unclear\,\cite{volovik88zetf123,chiu-16rmp35005}.

Motivated by the recent progress in the nanoscale design of artificial magnetic structures\,\cite{paris-experiment,WiesendangerPC,FrankePC} and the relevance of nanosocpic topological superconductors for quantum computing, we investigate the electronic 
properties of Shiba islands, \ie nanoscale islands of magnetic atoms placed on the surface of an $s$-wave superconductor. We identify key characteristic features of chiral Majorana modes localized along the edges of Shiba islands that can be probed via STS; these features provide direct insight into the topologically invariant Chern number of the system, allowing us to unambiguously distinguish topologically trivial from non-trivial phases. In particular, we find that (i) the differential tunneling conductance associated with the flow of charge from the STS tip into Majorana modes is quantized (apart from small finite-size effects) and proportional to $|\CC|$, and (ii) Majorana modes are equally spaced in energy, and carry a supercurrent whose chirality reflects the sign of the Chern number ${\rm sgn}(\CC)$. Moreover, we demonstrate that {\it zero-energy} Majorana modes -- crucial ingredients for topological braiding operations -- can be created by threading a $\pi$-flux through the island. These results represent crucial steps towards realizing topological quantum bits and designing complex topological phases.

\begin{figure*}[t!]
\centering
\includegraphics[width=15cm]{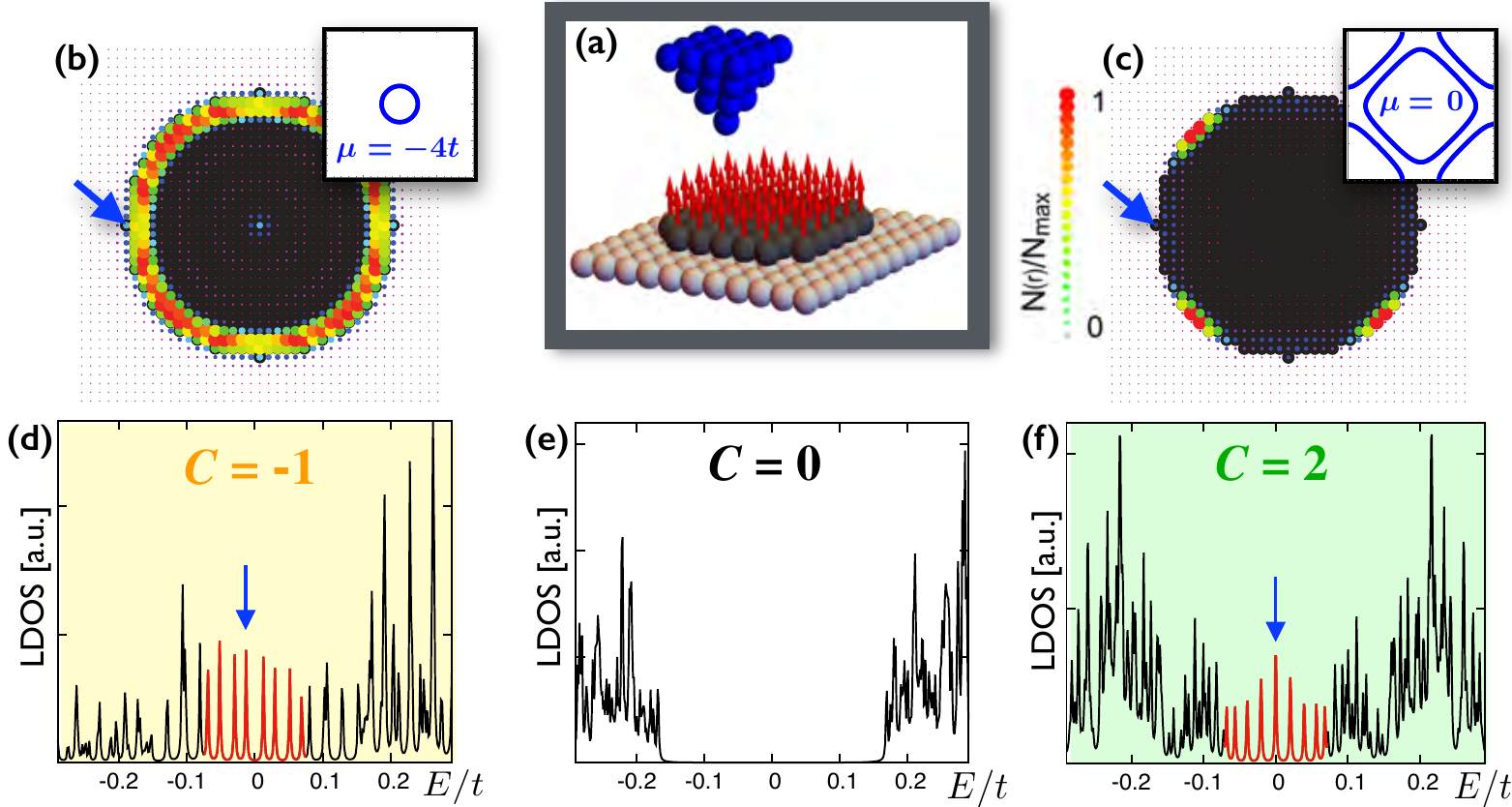}
\caption{
\textbf{Shiba island and LDOS:}
(a) Sketch of a Shiba island with STS tip.
The LDOS is shown for $\mu=-4t$ ($\CC=-1$) in (b,d), for $\mu=-2t$ ($\CC=0$) in (e), and for $\mu=0$ ($\CC=+2$) in (c,f).
Panels (d-f) display the energy-resolved spectra at the location marked by arrows in (b,c); equally spaced in-gap states are clearly visible in the topological phases in (d) and (f).
Panels (b,c) show the spatial weight distribution (color-coded) of the low-energy in-gap states marked by arrows in (d,f); the insets display the corresponding normal-state Fermi surfaces.
Parameters here and below are $(\alpha, \Delta_s, J)=(0.2, 0.3, 0.5)t$.
}
\label{fig:fig1}
\end{figure*}

%

\paragraph{Shiba-island model:}

To investigate the spectroscopic and transport properties of topological superconductors, we consider a finite two-dimensional island of magnetic adatoms deposited on the surface of an $s$-wave superconductor, as schematically shown in Fig.\,\ref{fig:fig1}(a). We take the magnetic moments of the adatoms to be aligned ferromagnetically, and describe them as classical moments, as spin-flip processes leading to Kondo screening are suppressed in a superconducting host. The system is then described by the Hamiltonian \,\cite{li-16nc12297} [assuming a square-lattice structure, as realized \eg in Pb(100)]
\begin{align}
H =& -t \sum_{{\langle \bf r r^\prime}\rangle, \sigma} \left( c_{{\bf r},\sigma}^\dag c^\pd_{{\bf r^\prime},\sigma}
+ {\rm H.c.}\right)  - \mu \sum_{{\bf r},\sigma} c_{{\bf r},\sigma}^\dag c_{{\bf r},\sigma}
\notag\\
+&\,i \alpha  \sum_{{\bf r},\sigma,\sigma'}\left( c_{{\bf r},\sigma}^\dag \sigma_{\sigma,\sigma'}^2 c^\pd_{{\bf r}+{\hat x},\sigma'} - c_{{\bf r},\sigma}^\dag \sigma_{\sigma,\sigma'}^1 c^\pd_{{\bf r}+ {\hat y},\sigma'}  + {\rm H.c.} \right) \notag\\
+& \,J \sum_{{\bf R},\sigma,\sigma'} c_{{\bf R},\sigma}^\dag \sigma_{\sigma\sigma'}^3 c^\pd_{{\bf R},\sigma'}  
+ \Delta_s \sum_{\bf r} \left( c_{{\bf r},\up}^\dag c^\dag_{{\bf r},\dw} + {\rm H.c.} \right)
\notag \\
-&t_{\rm tip} \sum_{\sigma} \left( c_{{\bf r},\sigma}^\dag d^\pd_{\sigma} + {\rm H.c.} \right)  + H_{\rm tip}\ ,
\label{ham-realspace}
\end{align}
where $c_{{\bf r},\sigma}^\dag$ creates an electron at lattice site ${\bf r}$ with spin $\sigma$, and $\sigma^i$ are the spin Pauli matrices. $t$ and $\Delta_s$ are the hopping and pairing amplitudes of the superconductor, $\mu$ is the chemical potential, $\alpha$ denotes the Rashba spin-orbit coupling arising from the breaking of the inversion symmetry at the surface\,\cite{nadj-perge-14s602}, $J$ is the magnetic exchange coupling, $t_{\rm tip}$ is the amplitude for electron tunneling from the STS tip into the system, and $H_{\rm tip}$ describes the electronic structure of the tip [see supplemental information (SI) Sec.\,I]. Finally, $\{{\bf R}\}$ denotes the sites of the magnetic Shiba island, and we consider only the two-dimensional surface of the superconductor.
For a Shiba lattice, i.e., when the surface is fully covered by magnetic adatoms\,\cite{li-16nc12297}, the system described by Eq.\,\eqref{ham-realspace} possesses
three topologically distinct phases with Chern numbers $\CC=-1,0$, and $+2$. One can tune between these phases by varying the chemical potential, with phase transitions at $\mu_{c,\pm}^{(1)} = \pm \sqrt{J^2-\Delta^2}$ and $\mu_{c,\pm}^{(2)} = \pm 4t \mp \sqrt{J^2-\Delta^2}$. Provided the superconductor is sufficiently thin, such tuning of $\mu$ can be achieved via gating\,\cite{privat}.
We
employ the non-equilibrium Keldysh Green's function formalism [see SI Sec.\,I for details]
to compute the local density of states (LDOS), the differential conductance, and charge currents in the Shiba islands.


\paragraph{Edge states:}
We begin by considering a Shiba island of radius $R=15 a_0$, and choose a set of parameters, $(\alpha, \Delta_s, J)=(0.2, 0.3, 0.5)t$ to ensure that the superconducting coherence length, $\xi \approx 6 a_0$ is smaller than $R$. In Fig.\,\ref{fig:fig1}\,(d,e,f), we plot the LDOS at the edge of the island for values of $\mu$ corresponding to the three different phases. The bulk-boundary correspondence\,\cite{qi-11rmp1057} implies that the island possesses low-energy edge states in the topological phases with $\CC=+2,-1$ [see Figs.\,\ref{fig:fig1}\,(d) and (f)], which are absent in the $\CC=0$ phase [Fig.\,\ref{fig:fig1}\,(e)]. Due to the finite size of the island, the topological edge modes are discrete levels, equally spaced in energy inside the superconducting gap. The spatially resolved LDOS of these low-energy mode reveals that most of their spectral weight is confined to the edge of the island [see Fig.~\ref{fig:fig1}(b,c)], thus confirming that these modes are indeed the putative edge modes  associated with the topological phases.
The distribution of the edge modes' spectral weight along the island's edge depends strongly on the shape of the underlying Fermi surface in the normal state: for a circular Fermi surface [$\CC=-1, \mu=-4t$, inset of Fig.~\ref{fig:fig1}(b)] the weight distribution is approximately uniform along the edge, while it is highly anisotropic for a nearly nested Fermi surface [$\CC=2, \mu=0$, inset of Fig.~\ref{fig:fig1}(c)].
With decreasing $\xi$ (see SI Sec.~II), the topological modes become more closely confined to the edges of the island, and their spectral weight in the interior of the island decreases. Similar edge states are also found in irregularly shaped islands (see SI Sec.~III), as their existence only relies on the topological character of the system's ground state, and not its particular geometry.

\begin{figure}[t!]
\centering
\includegraphics[width=8cm]{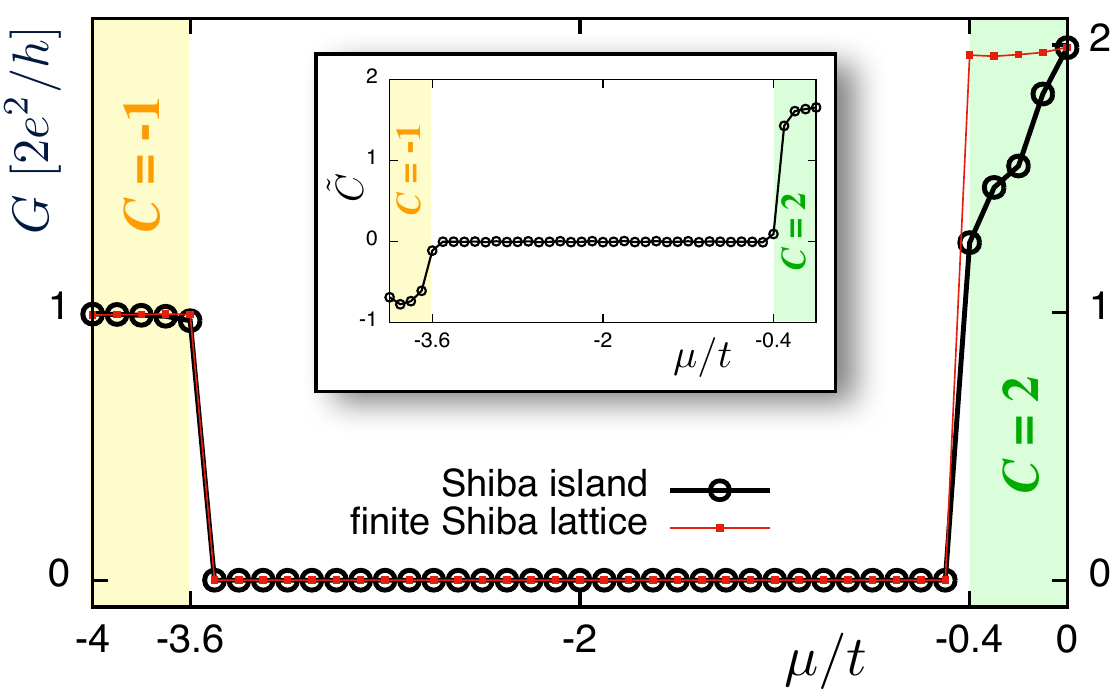} 
\caption{
\textbf{Conductance quantization:}
Differential conductance $G=dI/dV$ vs.\ chemical potential $\mu$ for the finite Shiba lattice (red) and the Shiba island (black), measured in a low-energy window, see text. $G$ not only distinguishes between the different phases but even is quantized to a remarkable accuracy as $G=\frac{2e^2}{h} |\CC|$. Here, $t_{\rm tip}=0.02t$. Inset: real-space Chern number normalized by coverage, $\tilde \CC$, demonstrating the topological character of the finite-size system, see SI for details.
}
\label{fig:fig2}
\end{figure}

\paragraph{Tunneling conductance:}
To further explore the properties of the topological edge modes, we consider the differential tunneling conductance associated with the flow of charges between an STS tip and a site ${\bf r}$ in the sample. As the main difference between the topological non-trivial and trivial phases lies in the presence of topological low-energy edge modes in the former, and their absence in the latter, we plot in Fig.\,\ref{fig:fig2} the maximum differential conductance $G= \max_{\Delta V}\left[ dI({\bf r})/d V \right]$ in a narrow voltage range, $\Delta V$, around zero voltage (see SI Sec.~IV). In the topological phases, $\Delta V$ contains only a single edge mode (being two-fold degenerate in the $\CC=2$ phase), with $G$ thus characterizing its conductance.

For a finite-sized Shiba lattice (see SI, Fig.~S2), we find the conductance of the topological edge modes to be quantized (within $0.1\%$) and given by $G=|\CC|\,2e^2/h $, with deviations from this quantized value becoming larger as one approaches the phase boundaries at $\mu_c^{(1,2)}$, see Fig.\,\ref{fig:fig2}.
This overall scale of $G$ is consistent with the observation that for a normal tip-superconductor junction, the quantum of conductance per fermionic degree of freedom is given by $2e^2/h$ due to Andreev scattering.
In the trivial $\CC=0$ phase, the conductance vanishes (for $t_{\rm tip} \rightarrow 0$) due to the absence of low-energy modes. The transition from $G \not = 0$ to $G = 0$ for the finite Shiba lattice occurs approximately at the same critical values of $\mu_c^{(1)} = \pm 0.4t, \mu_c^{(2)} = \pm 3.6t$ as in the thermodynamic limit. For the Shiba island with $R=15 a_0$, $G$ is still (nearly) quantized in the $\CC=-1$ phase, but deviations from the quantized value increase more rapidly in the $\CC=+2$ phase as $\mu_c^{(1)}$ is approached, which we attribute to the finite size of the island (the same conclusion also holds for irregularly shaped islands, see SI Sec.III).

To further substantiate the topological character of the nanoscopic system, we adapt and modify a real-space Chern number formalism previously used for disordered topological insulators\,\cite{prodan1} and apply it to our Shiba islands. Numerical results for the modified Chern number, $\tilde\CC$, are in the inset of Fig.\,\ref{fig:fig2}, for details see SI Sec.~V. While $\tilde\CC$ does not reach integer values for the nanoscopic island (as topological invariants are strictly defined only in the thermodynamic limit), it clearly distinguishes between topologically trivial and non-trivial phases; moreover it correctly reflects ${\rm sgn}(\CC)$ and also allows one to quantitatively  distinguish between the $\CC=-1$ and $\CC=2$ phases.

\begin{figure}[t!]
\centering
\includegraphics[scale=0.55]{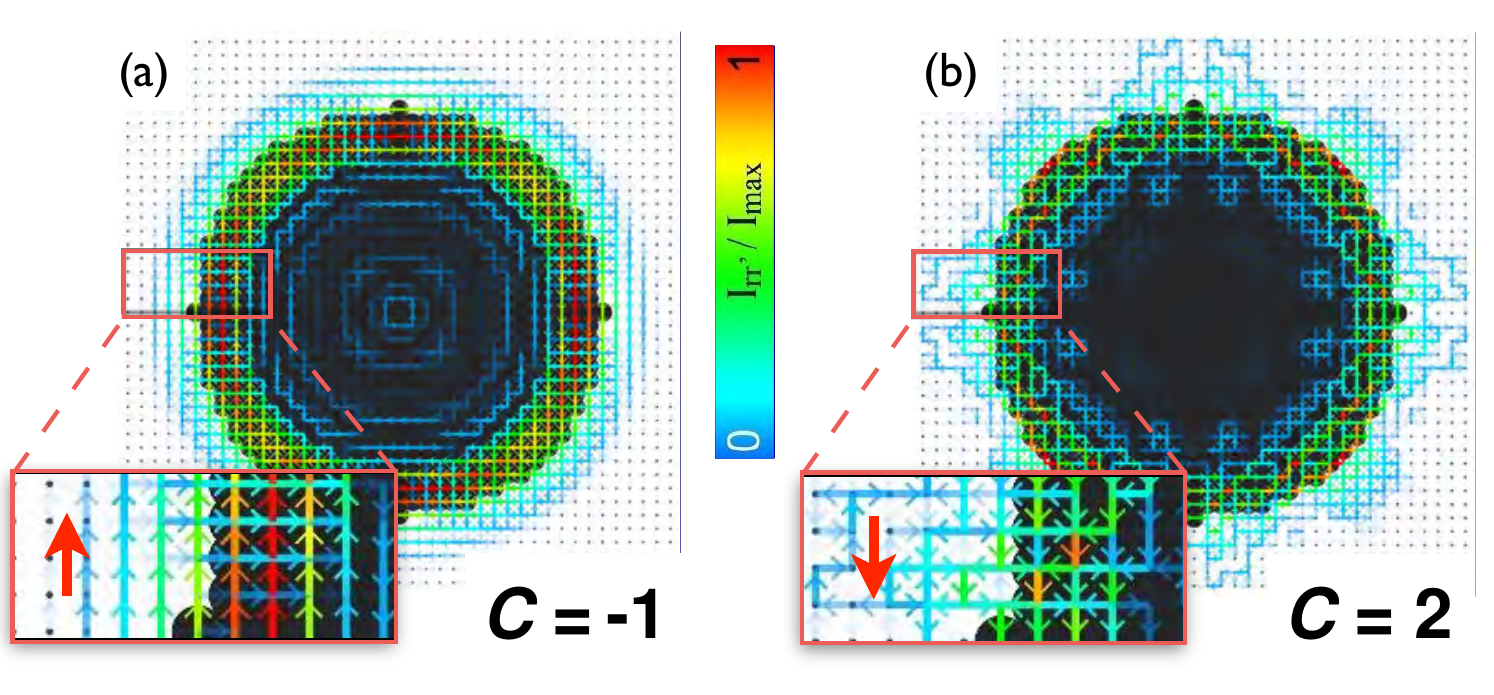}
\caption{
\textbf{Supercurrents:}
Spatial distribution of the supercurrents ($V\!=\!0$) carried by the lowest energy edge modes in the (a) $\CC=-1$ $(\mu\!=\!-4t)$, and (b) $\CC=+2$ $(\mu\!=\!0)$ phases.  The direction of the current flow depends on ${\rm sgn}(\CC)$ (see insets) with the currents flowing clockwise in (a) and counter-clockwise in (b).
}
\label{fig:fig4}
\end{figure}
\paragraph{Supercurrents:}
The combination of magnetic moments and Rashba interaction leads to the emergence of superconducting chiral spin-triplet correlations, whose nature reflects the broken time-reversal symmetry (TRS) inside the Shiba island, and the preserved TRS outside (see SI Sec.VI). Such chiral spin-triplet correlations imply the existence of a supercurrent in the system, \ie currents that flow even in the absence of any applied bias $V$. A plot of the supercurrents carried by the lowest-energy edge modes in the topologically non-trivial $\CC=-1$ and $\CC=2$ phases [Figs.\,\ref{fig:fig4}(a,b)] reveals that they are not only confined to the edge of the island, but that their chirality (i.e., direction of flow) reflects the sign of the Chern number, ${\rm sgn}(\CC)$. Recent advances in observing charge currents at small length scales\,\cite{Ji12,Rod16,Tet16} suggest that such supercurrents could be observed in the near future. This would imply that transport experiments could not only measure the magnitude of $\CC$ via the differential conductance, but also its sign, providing unprecedented insight into the nature of topological phases.

\begin{figure}[t!]
\centering
\includegraphics[scale=0.7]{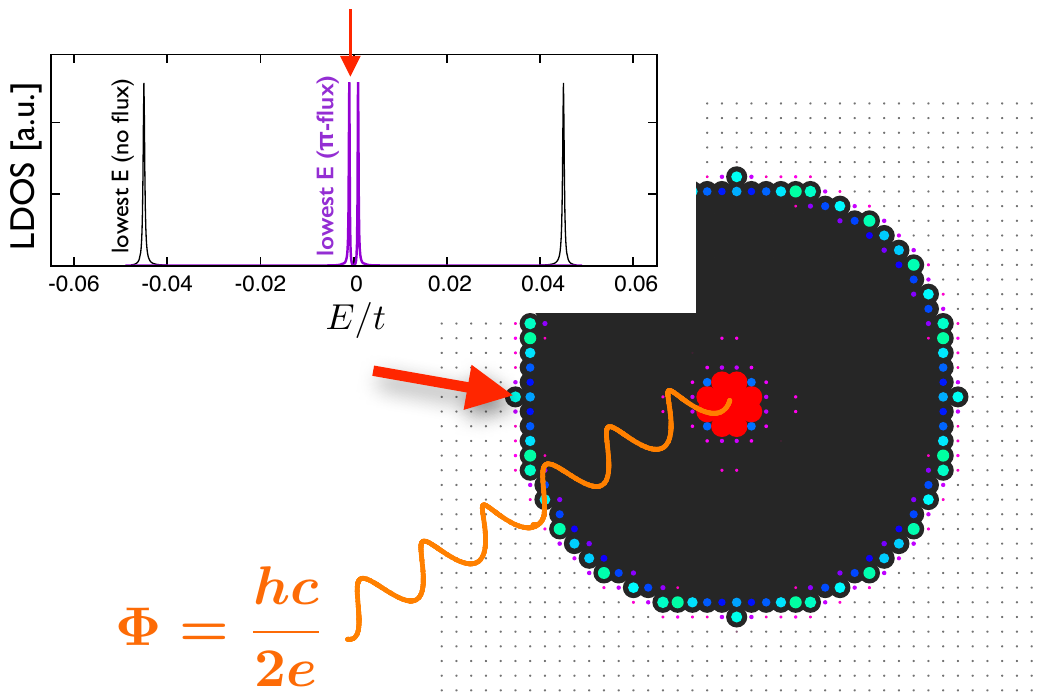}  
\caption{
\textbf{Creating Majorana zero modes:}
A $\pi$-flux threaded through the center of the Shiba island shifts states $E=\pm \epsilon/2$ (black) to almost zero energy (purple), with the remaining splitting being due to the small island size; the LDOS was measured at the location indicated by the red arrow. The spatial LDOS profile corresponds to one of the almost zero-energy state (arrow) and reveals that spectral weight is bound to the $\pi$-flux as well as to the island edge, reflecting the presence of two Majorana zero modes for large islands.
}
\label{fig:fig6}
\end{figure}

%
%
\paragraph{Majorana character of edge states:}
The realization of fault-tolerant topological quantum computing requires the creation of zero-energy Majorana modes with exotic braiding properties\,\cite{nayak-08rmp1083}.
Such Majorana zero modes have been predicted to exist \,\cite{kitaev01pu131,lutchyn-10prl077001,oreg-10prl177002,li-14prb235433} and subsequently been observed at the ends of the so-called Kitaev chains \cite{mourik-12s1003,nadj-perge-14s602,ruby-15prl197204,pawlak-16npj16035}. In contrast, the finite size of the Shiba islands considered above results in topological edge states that are located at discrete, and non-zero energies, $\pm E_i$. The symmetric states at $\pm E_i$ arise from the intrinsic particle-hole symmetry of the superconductor and can be considered a pair of Majorana fermions. To create an isolated zero-energy Majorana mode, it is necessary to spatially separate the two Majorana fermions forming a pair.
This can be achieved by inserting a $\pi$-flux in the center of a Shiba island: this flux creates a half-quantum vortex\,\cite{alicea-12rpp076501} and binds one of the two Majorana fermions. As a result, the Majorana modes become spatially separated, with one mode being located at the island's edge and one mode in the island's center at the $\pi$-flux, as follows from a plot of the low-energy LDOS in Fig.\,\ref{fig:fig6}. This spatial separation leads to an exponential suppression of the coupling between the Majorana modes with increasing radius of the island and a shift of their energies $\pm E_i$ to zero, as shown in the inset of Fig.\,\ref{fig:fig6}, where we contrast the low-energy part of the LDOS in the absence (black) and presence of the $\pi$-flux (purple). We thus obtain two zero-energy Majorana modes in the limit $R \rightarrow \infty$, which can be thought of as analogue to the Majorana bound states at the ends of a Kitaev or Shiba chain.

\paragraph{Summary:}
Identifying the characteristic electronic and transport properties of topological phases in nanoscopic Shiba islands is a key step in the quantum engineering of Majorana fermions. The robustness and universality of the results (see SI Secs.~II and III) is of crucial importance for applications.
Moreover, designing the magnetic structure or geometry of Shiba islands may enable creating more complex topological states and manipulating the spatial structure of Majorana modes. Both of these possibilities hold exciting new potential for realizing topological quantum computing.

%
%
\section{Methods}
To investigate the spectroscopic and charge transport properties of nanoscopic topological superconductors, we employ the non-equilibrium Keldysh formalism. This approach allows us to simultaneously compute the local density of states, the current flowing between an STS tip and the system, and the associated differential conductance, as well as the supercurrents flowing inside the topological superconductor. Furthermore, we extended a formalism to compute the Chern number in real space, allowing to us characterize the topological phase of Shiba islands.


\section{Acknowledgements}
The authors thank B. A.\ Bernevig, T.\ Neupert, A.\ Petrescu, E.\ Prodan, J.\ R\"ontynen, K.\ J.\ Franke, A.\ Kapitulnik, and R. Wiesendanger for discussions.
This work was supported by the DFG through SFB 1143, SPP 1666, and GRK 1621 (S.R.\ and M.V.) and by the U.\ S.\ Department of Energy, Office of Science, Basic Energy Sciences, under Award No.\ DE-FG02-05ER46225 (E.M., S.C., and D.K.M.).


\end{document}


\title{Supplemental Information for: \\
Quantized Charge Transport in Chiral Majorana Edge Modes}
\author{Stephan Rachel}
\affiliation{Institut f\"ur Theoretische Physik, Technische Universit\"at Dresden, 01062 Dresden, Germany}
\author{Eric Mascot}
\affiliation{University of Illinois at Chicago, Chicago, IL 60607, USA}
\author{Sagen Cocklin}
\affiliation{University of Illinois at Chicago, Chicago, IL 60607, USA}
\author{Matthias Vojta}
\affiliation{Institut f\"ur Theoretische Physik, Technische Universit\"at Dresden, 01062 Dresden, Germany}
\affiliation{Center for Transport and Devices of Emergent Materials, Technische Universit\"at Dresden,
01062 Dresden, Germany}
\author{Dirk K.\ Morr}
\affiliation{University of Illinois at Chicago, Chicago, IL 60607, USA}
\date{\today}

\maketitle
\tableofcontents

%
%
\section{Keldysh Formalism}
\label{sec:keldysh}

Our starting point for the study of Shiba islands on a square lattice is the Hamiltonian $H=H_{\rm SC}+ H_{\rm tip}$ \,\cite{li-16nc12297} where
\begin{align}
H_{\rm SC} =& -t \sum_{{\langle \bf r r^\prime}\rangle \sigma} \left( c_{{\bf r},\sigma}^\dag c^\pd_{{\bf r^\prime},\sigma} + {\rm H.c.}\right)
- \mu \sum_{{\bf r}\sigma} c_{{\bf r},\sigma}^\dag c_{{\bf r},\sigma}
+ \Delta_s \sum_{\bf r} \left( c_{{\bf r},\up}^\dag c^\dag_{{\bf r},\dw} + {\rm H.c.}\right) \notag\\
+&\,i \alpha  \sum_{{\bf r}\sigma\sigma'}\left( c_{{\bf r},\sigma}^\dag \sigma_{\sigma\sigma'}^2 c^\pd_{{\bf r}+{\hat x},\sigma'} - c_{{\bf r},\sigma}^\dag \sigma_{\sigma\sigma'}^1 c^\pd_{{\bf r}+ {\hat y},\sigma'} + {\rm H.c.} \right)  \notag\\
+& \,J \sum_{{\bf R}\sigma\sigma'} c_{{\bf R},\sigma}^\dag \sigma_{\sigma\sigma'}^3 c^\pd_{{\bf R},\sigma'} -t_{{\rm tip}} \sum_{\sigma} \left( c_{{\bf r},\sigma}^\dag d^\pd_{\sigma} + {\rm H.c.} \right)  \ .
\label{ham-realspace}
\end{align}
Here, $c_{{\bf r},\sigma}^\dag$ creates an electron at lattice site ${\bf r}$ with spin $\sigma$, and $\sigma^i$ are the spin Pauli matrices. $t$ and $\Delta_s$ are the hopping and pairing amplitudes of the superconductor, $\mu$ is the chemical potential, $\alpha$ denotes the Rashba spin-orbit coupling arising from the breaking of the inversion symmetry at the surface\,\cite{nadj-perge-14s602}, and $J$ is the magnetic exchange coupling. The lattice sites $\{{\bf R}\}$ correspond to the sites of the magnetic adatoms. Moreover, $t_{{\rm tip}}$ is the amplitude for electrons tunneling from the STS tip into the superconductor, with $d^\pd_{\sigma}$ annihilating an electron with spin $\sigma$ in the tip, and $H_{{\rm tip}}$ describes the electronic structure of the tip, as discussed below.

To investigate the transport properties of such magnetic islands, we employ the non-equilibrium Keldysh Green's function formalism \cite{keldysh65jetp1018,caroli-71jpc916}, which allows us to compute not only the tunneling current between the STS tip and the superconductor, but also the currents flowing on the surface of the $s$-wave superconductor. We note that due to the Rashba spin orbit coupling, the flow of charge between two sites in the system can be accompanied by a spin flip. Within the Keldysh formalism, the spin-resolved current flowing between sites $i$ and $j$ therefore has to be generalized to
\begin{equation}
I^{\sigma \sigma^\prime}_{ij} = -2\textit{g}_{\textit{s}}\dfrac{\textit{e}}{\hbar}\int_{-\infty}^{\infty} \dfrac{d\omega}{2\pi} {\rm Re} \left[ {\hat T}^{\sigma \sigma^\prime}_{ij} G^{<}(i,\sigma; j,\sigma^\prime; \omega)\right]  ,
\label{eq:I}
\end{equation}
where $G^{<}(i,\sigma; j,\sigma^\prime; \omega)$ is the full, normal lesser Green's function that describes the propagation of an electron with spin $\sigma$ at site $i$ to an electron with spin $\sigma^\prime$ at site $j$. Here, the matrix ${\hat T}^{\sigma \sigma^\prime}$ describes the hopping between two nearest-neighbor sites, which is either given by $-t$ if $\sigma =\sigma^\prime = \uparrow,\downarrow$, or by the Rashba coupling $\alpha$ if $\sigma \not =\sigma^\prime$.

To compute $G^{<}(i,\sigma; j,\sigma^\prime; \omega)$, we first define the Matsubara Green's function matrix in frequency space using the effective action
\begin{align}
S =& \frac{1}{\beta} \sum_{\omega_n >0} \Psi^\dagger(i \omega_n) \hat{G}^{-1}(i\omega_n) \Psi(i \omega_n)
\label{eq:eff_action}
\end{align}
where the spinor $\Psi^\dagger(i \omega_n)$ is defined via
\begin{align}
\Psi^\dagger(i \omega_n) = \left(d^\dagger_{\uparrow}(i \omega_n) ,d^\dagger_{\downarrow}(i \omega_n), d_{\downarrow}(-i \omega_n), d_{\uparrow}(-i \omega_n),  \ldots ,c^\dagger_{{\bf r},\uparrow}(i \omega_n) ,c^\dagger_{{\bf r},\downarrow}(i \omega_n), c_{{\bf r},\downarrow}(-i \omega_n),c_{{\bf r},\uparrow}(-i \omega_n), \ldots \right)
\end{align}
where ${\bf r}$ is a site in the superconductor. $\hat{G}(i\omega_n)$ is obtained from the Dyson equation
\begin{align}
\hat{G}(i\omega_n) = \left\{ \left[ \hat{g} (i\omega_n) \right]^{-1}  - \hat{H}_0 \right\}^{-1}.
\end{align}
Here, $\hat{H}_0$ is the Hamiltonian matrix defined using the Hamiltonian of Eq.(\ref{ham-realspace}) via
\begin{align}
H_{\rm SC} =& \frac{1}{2} \Psi^\dagger \hat{H}_0 \Psi
\label{eq:H_matrix}
\end{align}
with spinor
\begin{align}
\Psi^\dagger = \left(d^\dagger_{\uparrow}, d^\dagger_{\downarrow}, d_{\downarrow}, d_{\uparrow}, \ldots , c^\dagger_{{\bf r},\uparrow}, c^\dagger_{{\bf r},\downarrow}, c_{{\bf r},\downarrow}, c_{{\bf r},\uparrow}, \ldots \right) \ .
\end{align}
Note that the factor of $1/2$ in Eq.(\ref{eq:H_matrix}) arises since we consider particle- and hole-like operators for both spin-projections in the definition of $\hat{G}(i\omega_n)$ and the spinor $\Psi^\dagger(i \omega_n)$ in Eq.(\ref{eq:eff_action}). Finally, $\hat{g} (i\omega_n)$ is the Green's function matrix that represents decoupled and non-interacting sites in the system (see below).

To obtain the lesser Green's function in Eq.(\ref{eq:I}), we define lesser and retarded Green's function matrices $\hat{G}^{<,r}$ in real space whose $({\bf r r'})$  elements are given by $\hat{G}^{<,r}_{\bf r r'}$, and employ the Dyson equations in frequency space
\begin{subequations}
\begin{align}
\hat{G}^{<} &= \hat{G}^{r}\left[ \left(\hat{g}^{r} \right)^{-1} \hat{g}^{<}
\left( \hat{g}^{a} \right)^{-1} \right] \hat{G}^{a} \label{eq:fullGa} \\[5pt]
\hat{G}^{r} &=  \left[ \left( \hat{g}^{r} \right)^{-1}  - \hat{H}_0 \right]^{-1}
\label{eq:fullGb}
\end{align}
\end{subequations}
Here, $\hat{g}^{x}$ $(x=r,a,<)$ are given by
\begin{equation}
\hat{g}^{x}= \left(
\begin{array}{cc}
\hat{g}_{{\rm tip}}^{x} & 0  \\
0 & \hat{g}_{\rm SC}^{x}
\end{array}%
\right)
\end{equation}
where $\hat{g}_{\rm SC}^{x}$ and $\hat{g}_{{\rm tip}}^{x}$ are the Green's function matrices describing the $s$-wave superconductor and the Shiba island, and the tip, respectively. Moreover, $\hat{g}_{\rm SC}^{x}$ are diagonal matrices with elements
\begin{subequations}
\begin{align}
g_0^r(\omega) &= \frac{1}{\omega + i \delta } \\[5pt]
g_0^<(\omega) &= -2i n_F(\omega) {\rm Im}\{ g_0^r(\omega)\}
\end{align}
\end{subequations}
where $n_F(\omega)$ is the Fermi distribution function, in the superconductor. Moreover, $\hat{g}_{{\rm tip}}^{x}$ are diagonal matrices with elements
\begin{subequations}
\begin{align}
g_{{\rm tip}}^r(\omega) &= -i \pi \\[5pt]
g_{{\rm tip}}^<(\omega) &= -2i   n_F(\omega-eV) \ {\rm Im}\{ g_{{\rm tip}}^r(\omega)\}
\end{align}
\end{subequations}
implying that the tip's density of states is equal to unity and that we consider the wide band limit. Moreover, $e$ is the electron charge, and $V$ is the potential difference between the tip and the grounded superconductor. The spin-resolved local density of states, $N_\sigma({\bf r}, E)$ at site {\bf r} and energy $E$ is obtained from Eq.(\ref{eq:fullGb}) via
\begin{equation}
N_\sigma({\bf r}, E=\hbar \omega) = -\frac{1}{\pi} {\rm Im} \{ {\hat G}^r_{\bf rr}(\omega)\} \ .
\end{equation}

\section{Changing the Ratio between Coherence Length and System Size}

\begin{figure}[h]
\centering
\includegraphics[width=13cm]{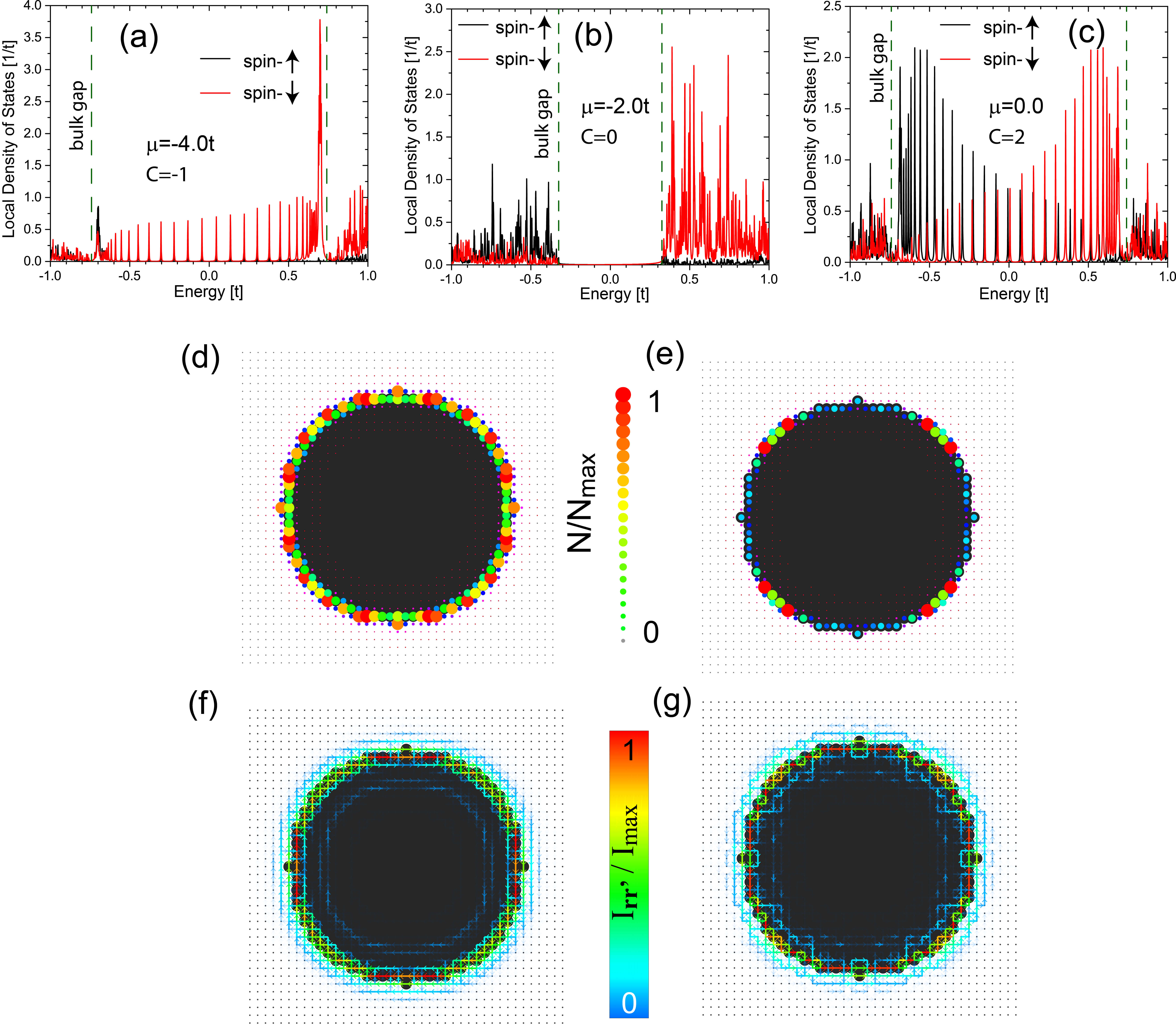}
\caption{(a) - (c) Spin-resolved LDOS at the edge of the Shiba island in the $\CC=-1$ $(\mu=-4t$), $\CC=0$ $(\mu=-2t$), and $\CC=2$ $(\mu=0$) phases, respectively, for a set of parameters $(\alpha, \Delta_s, J)=(0.8, 1.2, 2.0)\,t$. Total LDOS for the lowest-energy topological edge modes in the topological (d) $\CC=-1$ $(\mu=-4t$) and (e) $\CC=2$ $(\mu=0$) phases. Suppercurrents carried by the lowest-energy topological edge modes in the topological (f) $\CC=-1$ $(\mu=-4t$) and (g) $\CC=2$ $(\mu=0$) phases.}
\label{figSI-1}
\end{figure}

If the superconducting coherence length, $\xi$, is reduced in comparison to the system size, we expect that the edge modes are more strongly localized in the vicinity of the edges.
To investigate the effects of a shorter coherence length on the physical properties of Shiba islands, we consider a set of parameters $(\alpha, \Delta_s, J)=(0.8, 1.2, 2.0)\,t$ which yields a coherence length $\xi \approx 1.25 a_0$, which is about 5 times smaller than the one considered in the main text. In Figs.~\ref{figSI-1}(a)-(c), we present the resulting LDOS at the edge of the Shiba island in the $\CC=-1$ $(\mu=-4t$), $\CC=0$ $(\mu=-2t$), and $\CC=2$ $(\mu=0$) phases, respectively.
In Figs.~\ref{figSI-1}(d) and (e) we present the total LDOS for the lowest-energy topological edge modes in the topological $\CC=-1$ $(\mu=-4t$) and $\CC=2$ $(\mu=0$) phases. A comparison with Fig.1 in the main text reveals that due to the decreased superconducting coherence length, the edge modes are much more narrowly confined to the edge of the island. The same conclusion also holds for the spatial form of the supercurrents carried by the lowest-energy modes, shown in Figs.~\ref{figSI-1}(f) and (g). As discussed in the main text, we find that the chirality of the supercurrents is determined by the sign of the Chern number $\CC$. Independent of the ratio between $\xi$ and system size, we again find that the conductance is quantized in the topological phases.

\section{Irregularly Shaped Shiba Islands}
\label{sec:disorder}

\begin{figure}[t!]
\centering
\includegraphics[width=15cm]{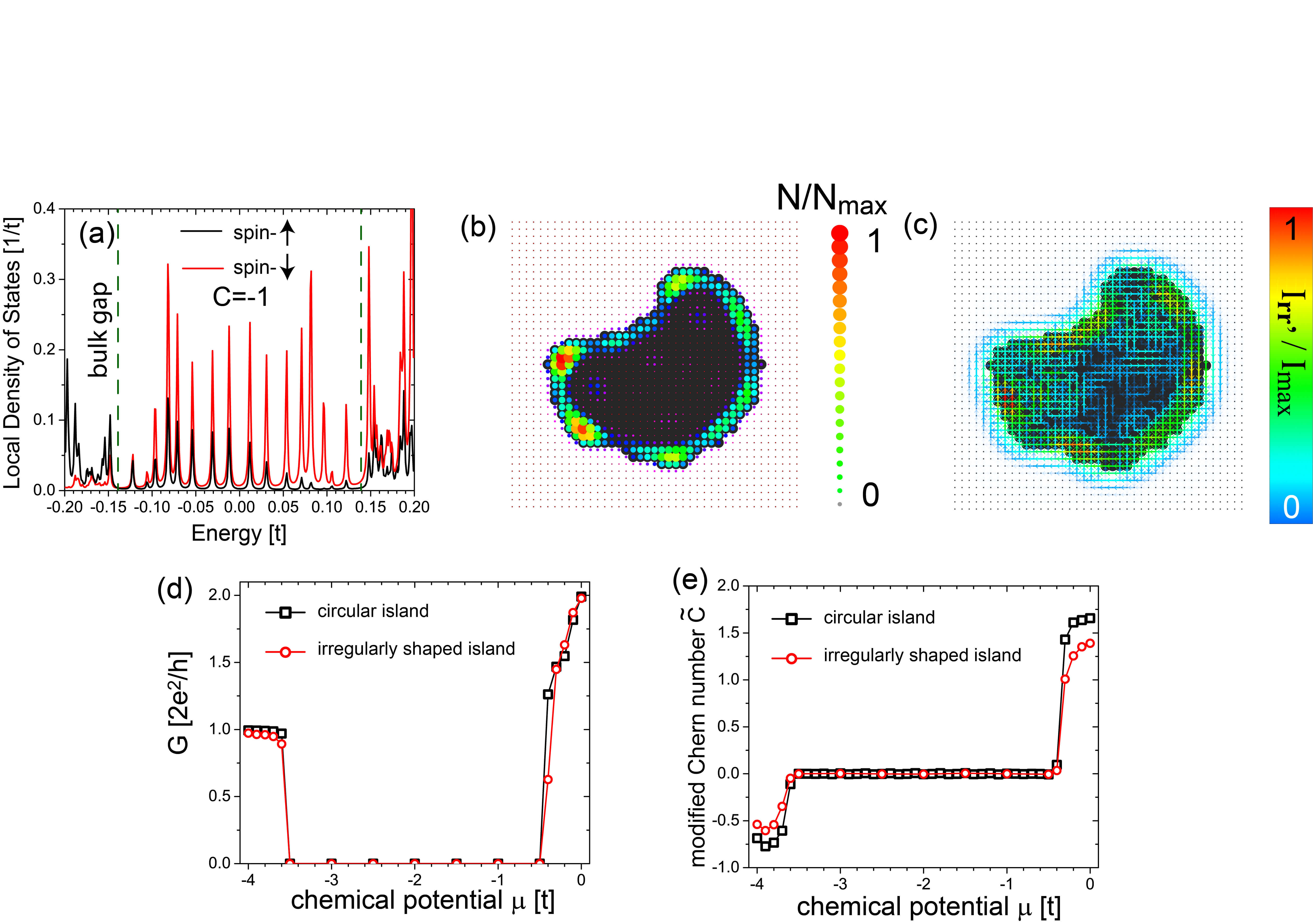}
\caption{Disordered Shiba island of magnetic adatoms with no rotational or mirror symmetries. (a) Energy-resolved LDOS in the $\CC=-1$ phase $(\mu=-4t)$. (b) Spatial LDOS of the lowest-energy edge mode shown in (a).  (c) Supercurrent carried by the lowest-energy edge mode shown in (a). (d) Differential conductance, $G$, and (e) modified Chern number ${\tilde \CC}$, as a function of $\mu$. Parameters used: $(\alpha, \Delta_s, J)=(0.2, 0.3, 0.5)\,t$.}
\label{figSI-7}
\end{figure}

In the main text, we considered a Shiba island that possesses the same spatial symmetries as the underlying lattice of the $s$-wave superconductor (\eg mirror and discrete rotational symmetries). Unless such highly ordered islands can be experimentally created using atomic manipulation techniques (as in the case of molecular  graphene\,\cite{moon-09nn167}), it is very likely that the experimental realization of Shiba islands will result in disordered or irregularly shaped islands. The question therefore arises to what extent the properties of the topological phases, such as their quantized conductance, are robust against deformations in the shape of the island (as long as the topological phase is not destroyed).

To investigate this question, we consider the irregularly-shaped magnetic island possessing no spatial symmetries shown in Fig.\,\ref{figSI-7}(b).
Despite its irregular shape, we find that the electronic and transport properties of the Shiba island, associated with the topological phases, remain qualitatively and to a large extent quantitatively unchanged. In Fig.~\ref{figSI-7}(a), we present the LDOS at the edge of the island in the $\CC=-1$ phase, which, as shown in Fig.~1 of the main text, exhibit a series of equally (in energy) spaced edge modes. A plot of the spatially resolved LDOS in Fig.~\ref{figSI-7}(b) for the lowest-energy edge mode in Fig.~\ref{figSI-7}(a) reveals that the edge mode is still strongly localized along the edge of the island, but penetrates further into the island due to its reduced size [cf. Fig.~1(b) of the main text]. The same conclusion also holds for the spatial form of the supercurrent shown in Fig.~\ref{figSI-7}(c) that is carried by the lowest-energy edge mode [cf. Fig.~3(b) of the main text]. The chirality of the supercurrents is the same as that for the circular Shiba island. Moreover, the differential conductance $G$ [Fig.~\ref{figSI-7}(d)] as well as the modified Chern number ${\tilde \CC}$ [Fig.~\ref{figSI-7}(e)], show very similar dependence on the chemical potential as those of the circular island. This reflects the persistent topological nature of the irregular-shaped island, as evidenced by a concomitant quantized tunneling conductance. Our results demonstrate the robustness of the topological phases and their intrinsic properties against deformations in the shape of the island.

\section{$\bs{I(V)}$ Curve and Differential Tunnel Conductance $\bs{G}$}

\begin{figure}[b!]
\centering
\includegraphics[width=8cm]{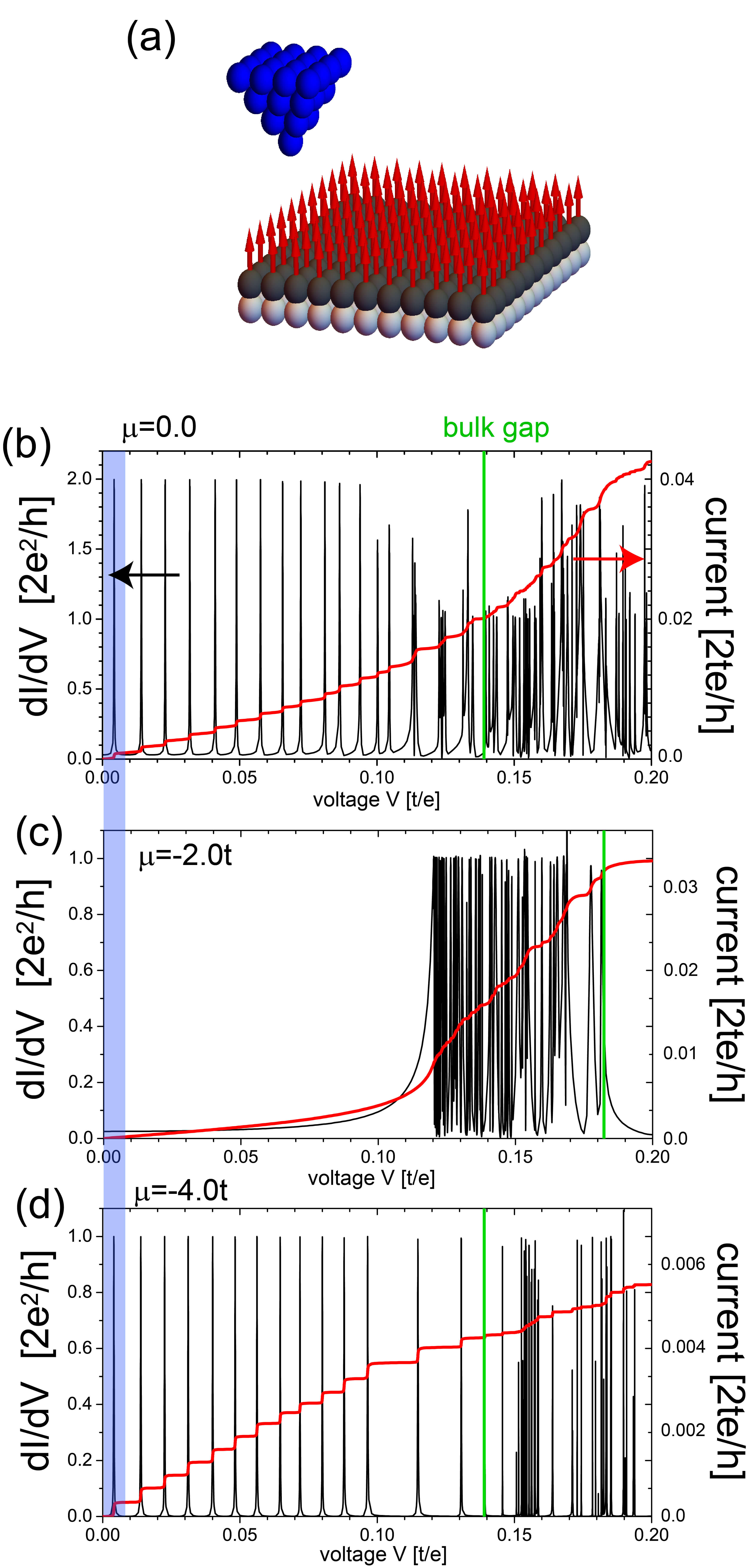}
\caption{(a) Schematic picture of a finite Shiba lattice with STS tip. Differential conductance, $G=dI/dV$, and current-voltage dependence $I(V)$ for a finite Shiba lattice of size $N_x=N_y=41$ in the (b) $\CC=2$ ($\mu=0$), (c) $\CC=0$ ($\mu=-2t$), and (d) $\CC=-1$ ($\mu=-4t$) phases. Here, $(\alpha, \Delta_s, J)=(0.2, 0.3, 0.5)\,t$ and $t_{{\rm tip}}=0.2t$. The light blue shaded area indicates the range of $\Delta E$. }
\label{figf:figSI-2}
\end{figure}
The main difference between the topological non-trivial ($\CC \not = 0$) and trivial ($\CC = 0$) phases lies in the presence of topological low-energy in-gap states (whose conductance is quantized) in the former, and the absence of these states in the latter. To characterize this difference, we consider the maximum differential conductance $G= \max_{\Delta V}\left[ dI({\bf r})/d V \right]$ in a narrow voltage range, $\Delta V$, around zero voltage (light blue area in Fig.~\ref{figf:figSI-2}). Here, ${\bf r}$ is the position in the superconductor where the electrons from the tip tunnel into. For concreteness, we consider a finite-sized Shiba lattice of size $N_x=N_y=41$. Moreover, we chose the range $\Delta V$ such that only a single edge mode with energy $E$ (which might be degenerate) lies within the energy window $0 \leq E \leq e \Delta V = \Delta E$, as shown in Figs.~\ref{figf:figSI-2}(b)-(d). We note that in the thermodynamic limit of a Shiba lattice, the $\CC=2$ phase possesses pairs of two degenerate edge modes. As a result, the conductance is twice as large as in the $\CC=-1$ phase [see Figs.~\ref{figf:figSI-2}(b) and (d), and Fig.~2 of the main text]. For any finite Shiba lattice or island, the degeneracy between the two modes is broken [see also the discussion in SI Sec.\,\ref{sec:ribbon}]. However, the non-zero electronic hopping $t_{{\rm tip}}$ between the STS tip and the superconductor leads to an energy broadening of the edge modes, such that for sufficiently large $t_{\rm tip}$ this broken degeneracy in the $\CC=2$ phase cannot any longer be resolved even for a finite-sized system,
as the width of the peaks is given by $\sim t_{\rm tip}^2$.
Thus the conductance is still twice as large as in the $\CC=-1$ phase [cf. Figs.~\ref{figf:figSI-2}(b) and (d)].
 In contrast, due to the absence of low energy edge modes in the topological trivial phases, no states lie within the energy window $\Delta E$, leading to $G \approx 0$. The small, but non-zero value of $G$ [see Fig.~\ref{figf:figSI-2}(c)] in the $\CC=0$ phase at low bias arises from the non-zero electronic hopping $t_{{\rm tip}}$, and vanishes in the limit $t_{{\rm tip}}\rightarrow 0$ .

%
%
\section{Real-Space Chern Number for Hybrid Systems}
\label{sec:Chern}
The topologically invariant Chern number $\CC$ is conventionally computed in momentum space for a translationally invariant system using\,\cite{thouless-82prl405}
\begin{equation}
\CC = \frac{1}{2\pi i} \int_{\rm BZ} {\rm tr}
\left\{ P_{\boldsymbol{k}} \left[ \partial_{k_1} P_{\boldsymbol{k}}, \partial_{k_2} P_{\boldsymbol{k}} \right] \right\} d\boldsymbol{k}
\label{eq:C_k}
\end{equation}
where
$P_{\bs{k}}$ is the $k$-decomposition of the projector $P$ onto the occupied states.
\begin{figure}[b!]
\centering
\includegraphics[width=9cm]{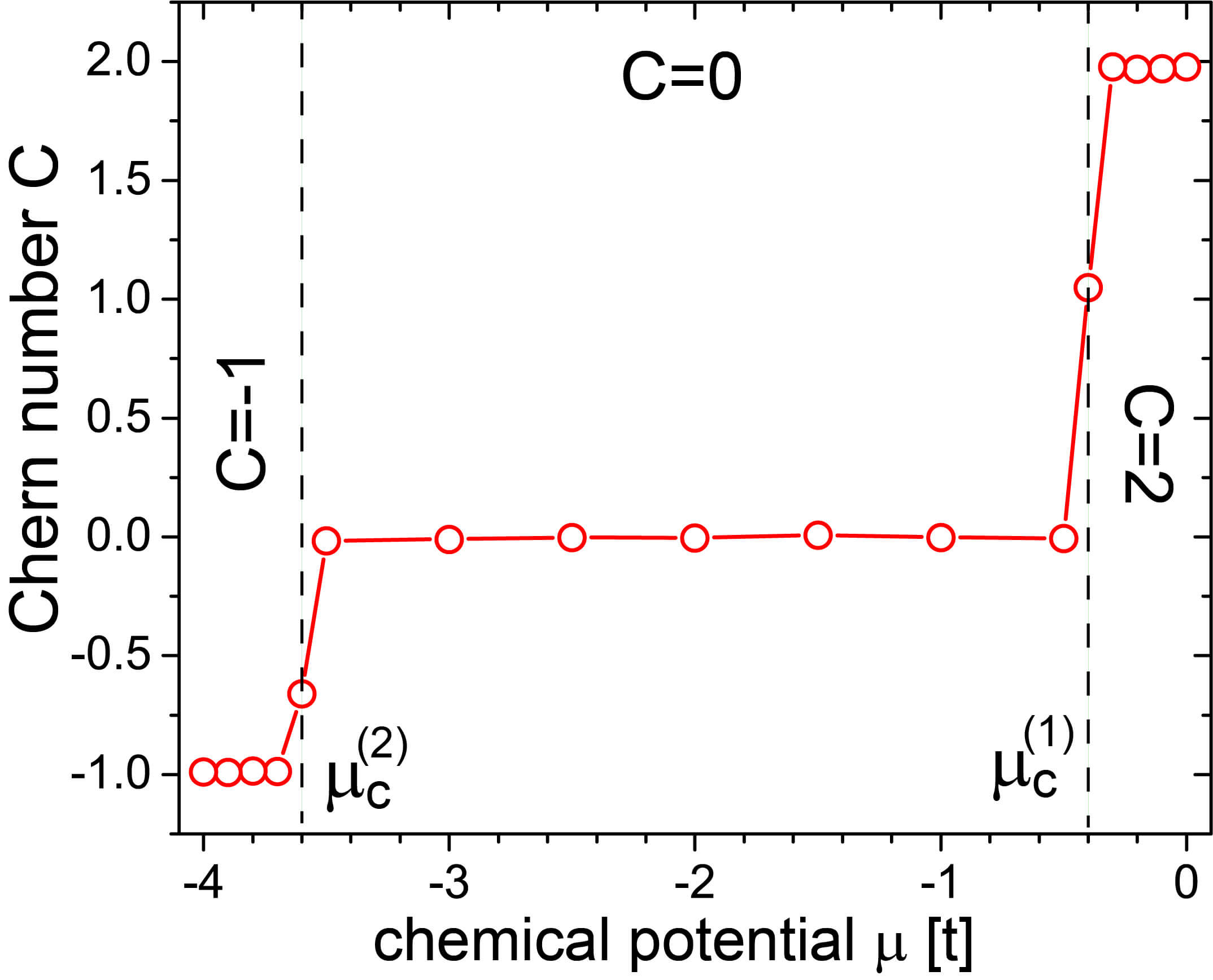}
\caption{Real-space Chern number (RSCN)  as a function of $\mu$ for a fully covered $41\times 41$ Shiba lattice with periodic boundary conditions. Parameters used: $(\alpha, \Delta_s, J)=(0.2, 0.3, 0.5)\,t$ leading to $\mu_c^{(1)} = -0.4\,t$ and $\mu_c^{(2)}=-3.6\,t$.}
\label{figSI-3}
\end{figure}

\begin{figure}[t!]
\centering
\includegraphics[width=12cm]{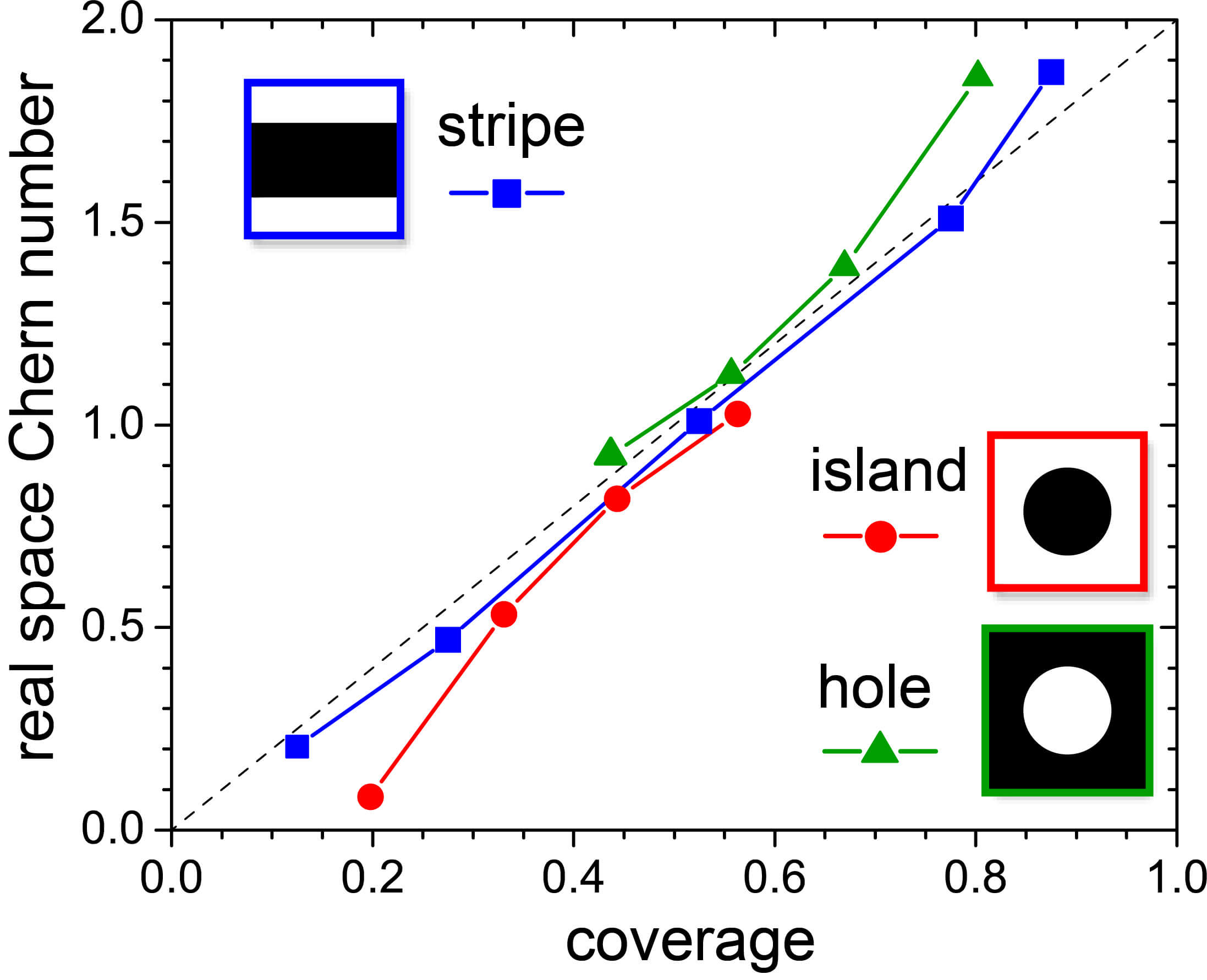}
\caption{Real-space Chern number (RSCN) as a function of coverage, \ie the ratio of sites covered by magnetic adatoms and the total number of sites, for $\mu=0$ corresponding to the $\CC=2$ phase. The lattice is a $41\times 41$ superconductor with periodic boundary conditions where a stripe of magnetic adatoms (blue) or an island of adatoms (red) is deposited. In addition, a covered system is considered where an island of adatoms is missing, yielding a hole (green).
Parameters used: $(\alpha, \Delta_s, J)=(0.8, 1.2, 2.0)\,t$.}
\label{figSI-4}
\end{figure}

\begin{figure}[t!]
\centering
\includegraphics[width=10cm]{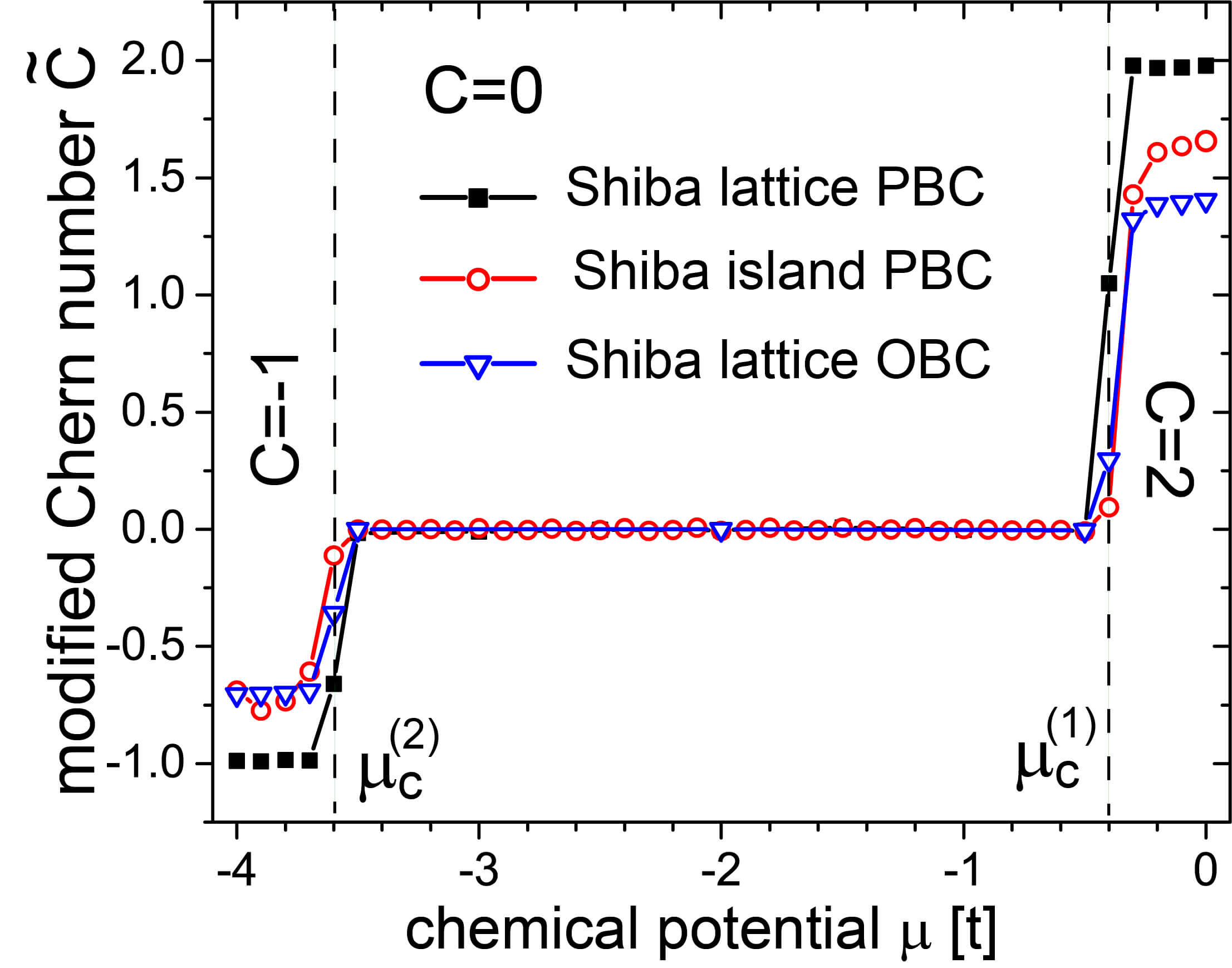}
\caption{Real-space Chern number divided by coverage, $\tilde \CC$, as a function of $\mu$
which reproduces the phase diagram of an infinitely large Shiba lattice\,\cite{li-16nc12297}. Results correspond to a Shiba island with a diameter of 30 atoms on a $41\times 41$ superconducting square lattice. In addition to the results for the Shiba island (red), we also present $\tilde \CC=\CC$ for the fully covered system (coverage = 1) with periodic boundary conditions (black) and open boundary conditions (blue).
Parameters used: $(\alpha, \Delta_s, J)=(0.2, 0.3, 0.5)\,t$.}
\label{figSI-5}
\end{figure}

The question of how this expression needs to be modified for systems with a broken translational invariance, for example due to the presence of disorder, was first considered by Bellissard {\it et al.}\,\cite{bellissard-94jmp5373}. They derived a formulation of the Chern number in real space, which for a translationally invariant system, and in the
thermodynamic limit, reproduces the results obtained from the momentum-space formulation in Eq.(\ref{eq:C_k}). Further progress was made by the pioneering work of Prodan {\it et al.}\,\cite{prodan-10prl115501,prodan11jpa113001,prodan17} who introduced an optimized real-space Chern number (RSCN), given by
\begin{align}
\CC =& 2\pi i
\sum_{\alpha} \langle 0,\alpha| P \left[-i [x_1,P], -i[x_2,P]\right]|0,\alpha\rangle
\label{eq:C_r}
\end{align}
where
\begin{align}
-i [x_i,P] = \sum_m c_m e^{-i m \boldsymbol{\Delta}_i \boldsymbol{x}} P e^{i m \boldsymbol{\Delta}_i \boldsymbol{x}}\ .
\end{align}
Within this formulation, the Chern number converges exponentially fast to the thermodynamic limit, such that one obtains very good approximations to the thermodynamic limit already for small real-space systems with periodic boundary conditions. To demonstrate this, we consider a $(41 \times 41)$ Shiba lattice with periodic boundary conditions (this implies that the surface of the $s$-wave superconductor is fully covered by magnetic atoms), and present in Fig.\,\ref{figSI-3} the resulting RSCN as a function of chemical potential $\mu$, for the same parameters as used in the main text. We see that the RSCN reproduces the expected quantization of the Chern number of the infinitely large system to high accuracy even for this rather small system size.
As the formulation of RSCN assumes periodic boundary conditions, the question naturally arises of whether it can be applied to Shiba lattices with open boundary conditions (OBC), or to finite-size Shiba islands on the surface of an $s$-wave superconductor, even if the latter possesses periodic boundary conditions.
In order to investigate the latter case, we plot in Fig.\,\ref{figSI-4} the Chern number for a Shiba stripe (``stripe'', see inset), a Shiba island (``island'') and a fully covered system where an island of magnetic adatoms is missing (``hole'') as a function of coverage (the coverage is defined as the ratio of sites covered by magnetic adatoms and the total number of sites in the system). We find that for such finite magnetic islands, the use of periodic or open boundary conditions shows very little quantitative effect on the RSCN.
For coverage between approximately 30 and 80 percent we obtain a linear dependence of the Chern number on coverage in all three cases. To understand this linear scaling, we note that the systems considered in Fig.~\ref{figSI-4} consist of a topologically non-trivial region (the magnetic Shiba island) with $\CC \not= 0$ and a trivial region (the surrounding superconductor) with $\CC=0$. Since the RSCN contains a summation over all lattice sites in the system (this can be seen when writing the projector 
$P_{\bs{k}}$
in real space), and not only a summation over sites that belong to the topological island, one can think of it as an averaged quantity: trivial regions yield a zero contribution to the Chern number, while non-trivial regions yield a finite contribution, resulting in the observed scaling of the RSCN with coverage of the non-trivial region. Therefore, to describe the topological nature of systems consisting of topologically trivial and non-trivial regions, we introduce a modified Chern number, $\tilde \CC$, defined as
\begin{equation}
\tilde \CC \equiv \frac{\CC}{{\rm coverage}}\ .
\end{equation}
The modified Chern number is not a topological invariant in a strict sense, as it does not reach integer values as expected from an invariant. Nevertheless, the modified Chern number provides important insight into the topological phase of the Shiba islands, as follows from a plot of $\tilde \CC$ as a function of $\mu$ shown in Fig.\,\ref{figSI-5} for the Shiba island considered in the main text. This plot demonstrates that $\tilde \CC$ (despite not being a strict topological invariant) retains features of a topological invariant: (i) it clearly distinguishes between phases with $\CC \not= 0$ and  $\CC=0$. In particular, for those values of $\mu$ where for the Shiba lattice with PBC one finds $\CC=0$, the Shiba island also possesses a RSCN that is strictly zero. (ii) The sign of $\tilde \CC$ is in all cases in agreement with the PBC results. (iii) Even quantitatively $\tilde \CC$ leads to reasonable results: for instance, for the Shiba island which would correspond to $\CC=-1$ ($\CC=2$) in the thermodynamic limit, we find $\tilde \CC \approx -0.8$ ($\tilde \CC \approx 1.7$). We therefore conclude that the modified Chern number $\tilde \CC$ is a valuable tool for the detection of topological phases.

Lastly, we consider the effects of a finite-size Shiba lattices with open boundary conditions (in contrast to the periodic boundary conditions considered above). Here, we find that switching from periodic to open boundary conditions (while keeping the coverage at unity) leads to a suppression of the Chern number by approximately $25 \%$ for the considered system sizes and for all values of $\mu$, as shown in Fig.~\ref{figSI-5}. As mentioned before, for Shiba islands the differece between periodic and open boundary conditions is negligible.

\section{Induced Spin-Triplet Correlations in the $\bs{p}$-wave Channel}

\begin{figure}[t]
\centering
\includegraphics[width=10cm]{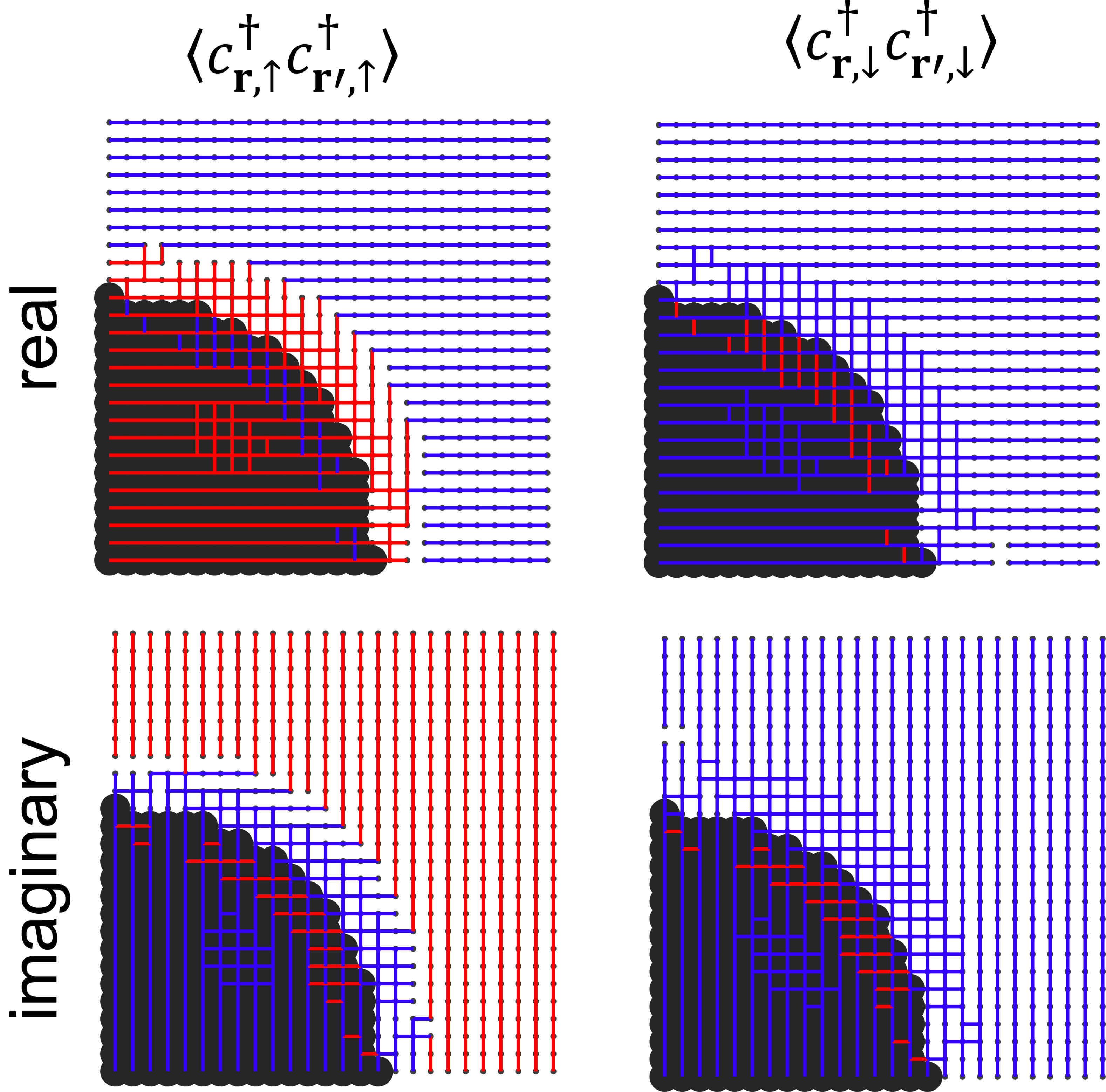}
\caption{Induced superconducting triplet correlations $\langle c^\dagger_{{\bf r},\sigma} c^\dagger_{{\bf r^\prime},\sigma} \rangle$ for $\sigma =\uparrow, \downarrow$. The upper row shows the sign of the real part, the lower row the sign of the imaginary part of the correlations -- positive (negative) sign is show in blue (red). The parameters are $(\alpha, \Delta_s, J) = (0.2, 0.3, 0.5)\,t$, and $\mu=-0.3t$ corresponding to the $\CC=2$ phase. Shown are just quarters of the Shiba island with radius $R=15a_0$.}
\label{fig:fig3}
\end{figure}
%

The combination of magnetic impurities, Rashba spin-orbit interaction, and $s$-wave superconductivity gives rise to the emergence of superconducting triplet correlations \cite{lutchyn-10prl077001,oreg-10prl177002,li-14prb235433}. To investigate the spatial form of these correlations, we consider the spin-triplet, equal-spin correlations on nearest-neighbor sites ${\bf r},{\bf r^\prime}$, as described by the correlation function $\langle c^\dagger_{{\bf r},\sigma} c^\dagger_{{\bf r^\prime},\sigma} \rangle$. These correlations describe superconducting pairing in the $p$-wave channel. In Fig.~\ref{fig:fig3} we present the sign of the real and imaginary parts of $\langle c^\dagger_{{\bf r},\sigma} c^\dagger_{{\bf r^\prime},\sigma} \rangle$ for $\sigma =\uparrow, \downarrow$ and $\mu=-0.3t$ inside the $\CC=2$ phase. The correlations are predominantly real along the horizontal links, and imaginary along the vertical links, suggesting a $\pm p_x \pm i p_y$ orbital structure of the induced triplet correlations. A closer analysis of the relative sign of the real and imaginary parts reveals that the correlations inside the droplet are $-p_x + i p_y$ for $\langle c^\dagger_{{\bf r},\uparrow} c^\dagger_{{\bf r^\prime},\uparrow} \rangle$ and $p_x + i p_y$ for $\langle c^\dagger_{{\bf r},\downarrow} c^\dagger_{{\bf r^\prime},\downarrow} \rangle$, which reflect, as expected, the broken time-reversal symmetry of the system due to the presence of magnetic defects. However, outside the magnetic island, the correlations are $p_x - i p_y$ for $\langle c^\dagger_{{\bf r},\uparrow} c^\dagger_{{\bf r^\prime},\uparrow} \rangle$ and $p_x + i p_y$ for $\langle c^\dagger_{{\bf r},\downarrow} c^\dagger_{{\bf r^\prime},\downarrow} \rangle$, which preserve the system's time-reversal symmetry. This implies that the nature of the induced triplet correlations changes between the interior and exterior of the droplet. We find that these relations between the signs of the induced triplet correlations hold both for the topologically trivial and non-trivial phases, despite the fact that in the topologically trivial phase there are no edge modes. The $\CC=0$ phase therefore represents an example of a system that exhibits superconducting (chiral) triplet ($p$-wave) correlations, but no corresponding edge modes.


%
%
\section{Results for Ribbon Geometry}
\label{sec:ribbon}

One of the main objectives of this article is to predict which topological properties of Shiba lattices will persist down to small nanoscopic Shiba islands.
To extrapolate between Shiba lattices with periodic boundary conditions (such as the one discussed in Fig.~\ref{figSI-3}) and finite-size Shiba islands with open boundary conditions, it is instructive to consider Shiba nano-ribbons -- systems with cylinder geometry --  which can be thought of a system which has PBC imposed along one and OBC imposed along another direction. Formally, one performs a Fourier transformation along the $x$ direction but remains in real space regarding the $y$ coordinate. Energy spectra can then be plotted with respect to the  momentum quantum number $k_x \equiv k$. Moreover, due to OBC along the $y$ direction the system possesses edges and carries thus $\CC$ edge modes (due to bulk-boundary correspondence), which can be studied as a function of $k$. Note that a ribbon carries $2\CC$ edge modes, $\CC$ per edge, while a system with OBC in $x$ and $y$ directions such as the Shiba island possesses only $\CC$ edge modes.

\begin{figure}[t!]
\centering
\includegraphics[scale=1.05]{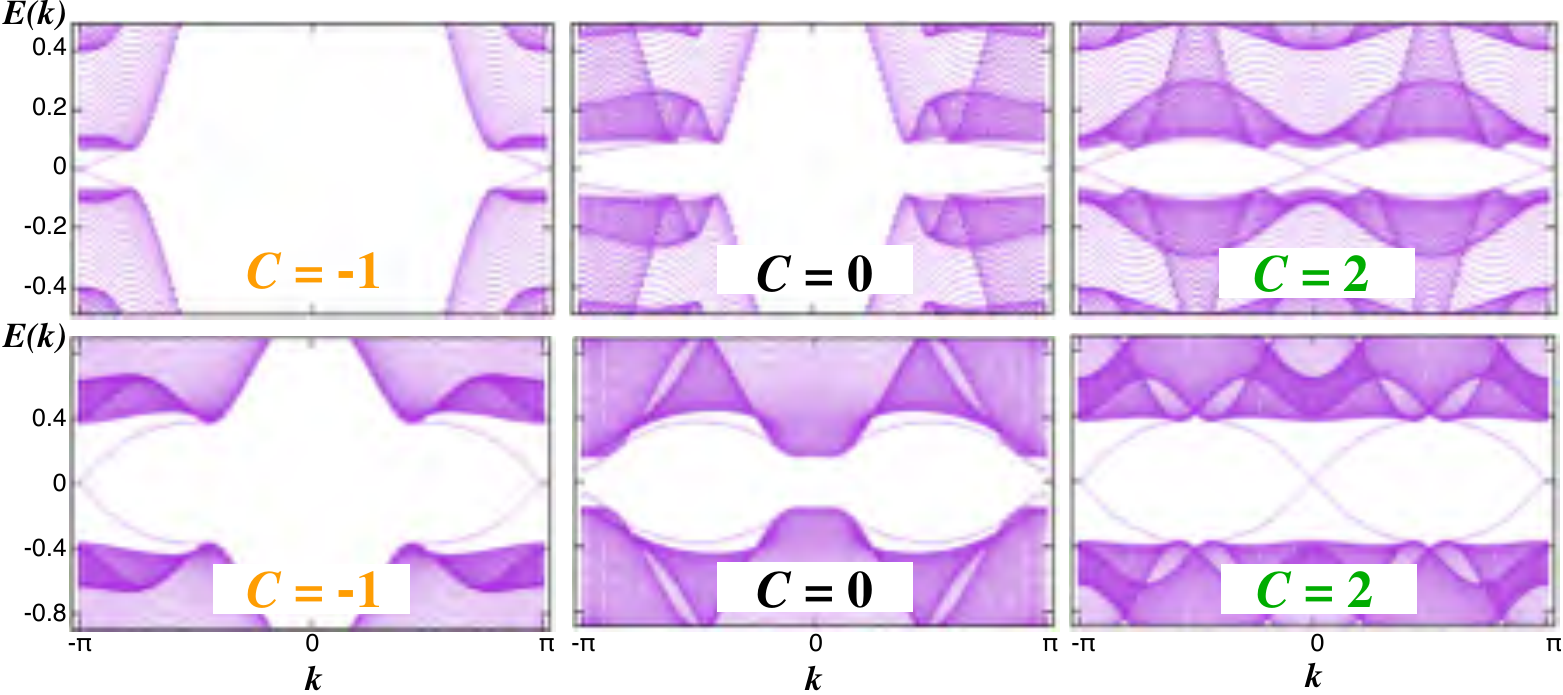}
\caption{Cylinder spectra on a ribbon consisting of 100 unit cells. Top row: $(\alpha, \Delta_s, J)=(0.2, 0.3, 0.5)\,t$. Bottom row: $(\alpha, \Delta_s, J)=(0.8, 1.2, 2.0)\,t$. Systems with Chern numbers $\CC=-1$, $0$, and $2$ correspond to $\mu/t=-4$, $-2$, and $0$, respectively.}
\label{figSI-cyl}
\end{figure}

The relation between the cylinder spectra shown in Fig.\,\ref{figSI-cyl} for two different parameter sets and different topological phases and the corresponding Shiba island LDOS plots [Fig.\,1 of the main paper] is obvious, as one simply has to project all energy levels at different $k$ values onto each other in order to obtain the global (\ie spatially integrated) DOS. The number of energy levels or energy peaks in the LDOS depends on the number of lattice sites. For the cylinder spectra $k$ is a free parameter and we can choose arbitrary discretizations, \eg the spectra in Fig.\,\ref{figSI-cyl} are shown for 150 $k$ values. Fig.\,\ref{figSI-cyl} discloses an interesting detail: for the $\CC=2$ phase, the two chiral edge modes are at different wave vectors. Of course the distinction via wavevectors becomes useless for the Shiba islands, but we see that the naive picture that the $\CC>1$ chiral edge modes are like identical copies on top of each other is by no means justified. For the LDOS plots of the Shiba island we should keep in mind, that energy levels do not necessarily need to come in pairs for $\CC=2$ and that the two dispersive Majorana modes might behave differently, in particular when the island shape is not symmetric (see Sec.\,\ref{sec:disorder}) or dirt and imperfections are present such as in realistic situations.

\bibliography{TSCtransport}